# 4f-hybridization effect on the magnetism of $Nd_2PdSi_3$

K. Mukherjee, Tathamay Basu, Kartik K. Iyer, and E.V. Sampathkumaran*

Tata Institute of Fundamental Research, Homi Bhabha Road, Colaba, Mumbai-400005, India

Among the members of the series $R_2PdSi_3$ ($R$= Rare-earths), the magnetic behavior of Nd compound is interesting in some respects. This compound is considered to order ferromagnetically (below 16 K), unlike other members of this series which order antiferromagnetically. In addition, magnetic ordering temperature ($T_o$) is significantly enhanced with respect to de Gennes scaled value. In order to understand the magnetism of this compound better, we have investigated the magnetic behavior in detail (under external pressure as well) and also of its solid solutions based on substitutions at Nd and at Si sites, viz., on the series, $Nd_{2-x}(Y, La)_xPdSi_{3-y}Ge_y$ (x, y = 0.2, 0.4, 0.8, and 1.2) by bulk measurements. The results overall establish that $Nd_2PdSi_3$ orders ferromagnetically below 16 K, but antiferromagnetic component seems to set in at very low temperatures. Notably, there is a significant suppression of $T_0$ for Y and Ge substitutions, compared to La substitution, for a given magnitude of unit-cell volume change, however qualitatively correlating with the separation between the layers of Nd and Pd-Si(Ge). On the basis of this observation, we conclude that 4f(Nd) hybridization plays a major role on the magnetism of the former solid solutions. To our knowledge, this work serves as a rare demonstration of 4f-hybridization effects on the magnetism of an Nd-based intermetallic compound.

## I. INTRODUCTION

It is a well-known fact that the 4f-orbital gets gradually localized as the 4f-occupation number is increased for rare-earths (*R*). The investigation of the consequences of partial extension of 4f-orbital on the solid state properties of *R*-intermetallics has been at the centre stage of modern condensed matter physics research. The fact however remains that such studies demonstrating 4f-hybridization effects have been to a large extent restricted to Ce compounds and to a much lesser extent to Pr compounds [1], but very little detailed work exists in the literature for Nd compounds barring some photoemission studies [2]. It is therefore of interest to focus investigations on Nd-based compounds.

In this article, we bring out interesting 4f-hybridization effects on the magnetism of the compound, $Nd_2PdSi_3$, belonging to a $AlB_2$-derived ternary family [3] (hexagonal structure, space group P6/mmm) with many exotic properties [4-21]. This ternary derivative arises out of ordered replacement of B-site by Pd and Si ions, resulting in two different chemical environments for *R* ions. The unit-cell is doubled within the basal plane with respect to $AlB_2$ unit-cell. The Pd and Si atoms form two-dimensional hexagonal rings and these layers are separated by *R* ions, as described in Ref. 4. This compound appears to be an interesting one in this family in the following respects. It is generally believed that this compound orders ferromagnetically [3, 7, 21] with a weak frequency-dependent ac susceptibility ($\chi$) features below $T_0$ [15], whereas in many other members of this series there is a paramagnetic-antiferromagnetic transition followed by spin-glass-like features, with lowering temperature. In addition, the observed magnetic ordering temperature ($T_o$= 16 K) is about 7 times larger than that expected on the basis of de Gennes scaling and only marginally lower than that observed [5] for $Gd_2PdSi_3$, $T_N$= 21 K. Therefore, in order to understand this compound better, a detailed study involving magnetization (*M*), heat capacity (*C*) and electrical resistivity ($\rho$) measurements was carried out. In addition, the same investigations were carried out on its solid solutions, based on substitutions at Nd site or at Si site, viz., on the series, $Nd_{2-x}(Y, La)_xPdSi_{3-y}Ge_y$ (x, y = 0.2, 0.4, 0.8, and 1.2), to shed more light about the behavior of Nd-4f in this chemical environment. In addition, high pressure magnetization studies on the parent compound were also carried out. The results, apart from revealing complex nature of the magnetism of this compound, bring out 4f-hybridization effect on the magnetism of Nd.

## II. EXPERIMENTAL DETAILS

Polycrystalline alloys of the solid solutions, $Nd_{2-x}(Y, La)_xPdSi_{3-y}Ge_y$ (x, y = 0.2, 0.4, 0.8, and 1.2), were prepared by arc-melting stoichiometric amounts of respective high-purity elements in an atmosphere of argon, and were subsequently homogenized by vacuum annealing in sealed quartz tube at 750°C for a week. X-ray diffraction studies established that all the specimens are single phase (Fig 1) within the detection limit (<2%) of this technique, forming in the proper structure. A comparison of x-ray diffraction patterns of the parent and doped compounds (inset of Fig 1) reveals a gradual shift of diffraction lines to the higher angle side for *Y*-series and to lower angle side for La/Ge-series, thereby establishing that the dopants go to respective sites. The lattice parameters of the compound are listed in Table 1.

The *dc M* in the temperature (*T*) range of 1.8-300 K for all specimens was carried out with the help of a commercial superconducting quantum interference device (Quantum Design) magnetometer. The *ac* susceptibility studies were also carried at selected frequencies. We have performed heat capacity measurements employing a physical property measurements system (PPMS, Quantum Design) as a function of $T$ in the range of 1.8-100 K. The $\rho$ measurements by standard four probe method, in the absence/presence of magnetic fields (upto $H$=100 kOe, $T$=1.8–300K), were performed with the same PPMS. A conducting silver paint was used for making electrical contacts of the leads with the samples. Further, $M$ measurements, were carried out employing a commercial cell (Easy-Laboratory Technologies Ltd, U.K.) in a hydrostatic pressure medium (upto ≤10 kbar at 4.2 K) of Daphne oil with the help of the above-mentioned SQUID magnetometer. Unless stated otherwise, all the measurements were done under zero-field cooled (ZFC) conditions of the specimens.

## III. RESULTS AND DISCUSSION
### A. Complex nature of the magnetism of $Nd_2PdSi_3$

Inverse of magnetic susceptibility ($\chi$) measured in a magnetic field of 5 kOe is plotted as a function of temperature [Fig. 2a] and the Curie Weiss-fit [$M/H=C/(T-\theta_p)$, where C is a constant] of the curve yields Curie-Weiss temperature ($\theta_p$) of ~ 8 K. The experimentally determined value of the effective moment of $3.5\mu_B/Nd^{3+}$ compares well with free $Nd^{3+}$ value of 3.6 $\mu_B$, while the positive value of $\theta_p$ points towards the dominance of FM interaction. However, the value of $\theta_p$ being less than $T_0$ could indicate the presence of AFM correlations competing with FM interactions in the paramagnetic state, as observed for $Tb_2PdSi_3$ in the same family [4]. In fact, the results on single crystals [21] revealed that the sign of $\theta_p$ (about 9.5 K) is positive for $H//$[001], whereas it is negative (about -2.3 K) along $H//$[100], thereby offering evidence for anisotropic nature of the magnetic interaction; interestingly, the difference between the magnitudes turns to be nearly the same as the value of $\theta_p$ in the polycrystalline sample.

The temperature dependence of $dc$ magnetization under ZFC and field-cooled (FC) condition at 100 Oe at low temperatures is shown in figure 2b. The features attributable to magnetic ordering are clearly seen in the data near 15.5 K. A peak is observed just below $T_0$ in the ZFC curve, which is followed by a drastic fall in the magnetization value, but the FC curve continuously rises with decreasing temperature. This fall in ZFC curve is generally not characteristic of long-ranged ordered FM systems, in which the magnetization below $T_o$ varies as inverse of demagnetization factor and is expected to be constant [22]. However, in the case of ferromagnetic systems, one can see a bifurcation of ZFC-FC curves if the coercive field becomes larger than the externally applied field. We find that the coercive field in our sample diminishes with increasing temperature, e.g., 1.1 and 0.19 kOe at 1.8 and 5 K respectively. It may be added that such a bifurcation is also sometimes observed [23] in anisotropic systems.

If we look at the $M(H)$ curve at 1.8 K [Fig. 2c], a magnetic hysteresis (which is characteristic of FM systems) is observed and the magnetization does not saturate till high fields. The studies extended with a commercial vibration sample magnetometer (Quantum Design) to higher fields revealed that this trend persists till the highest field employed (160 kOe). Also in Ref. 21, the isothermal magnetization was reported up to 120 kOe and there is a weak variation for easy axis $H//$[001] till high fields. The non-saturation tendency of magnetization is usually noted for systems with finite AFM correlations even at high fields at 1.8 K. A point of emphasis is that the virgin curve at low fields is nearly flat, resulting in S-shaped curvature. With increasing temperature, this S-shaped curvature is gradually suppressed [Fig. 2d]. One would naively expect that this step arises from anisotropic coercive ferromagnetism. On the other hand, such a flatness can be observed if AFM correlations are present in the virgin state, undergoing metamagnetic transition with increasing field. Therefore it is not easy to distinguish between these two possibilities. However, a well-known signature of a disorder-broadened first-order transition (as a function of *H*) is that the virgin state tends to lie outside the envelope curve (in the plot of *M* versus *H*). A careful look at the inset of figure 6b in Ref. 21 suggests that, at 2.5 K, the virgin curve beyond the flat-region for *H*//[001] superimposes over the increasing leg of the envelope curve, which is not typical of a ferromagnetic loop. In fact, we obtained a single crystal from the authors of Ref. 21 and measured *M(H)* at 1.8 and 5 K for *H*//[110] (see figures 2e and 2f). We find that the virgin curve (marked by the path 1 in the figure) clearly lies outside the envelope curve when the magnetic-field is applied along a basal plane at 1.8 K. The virgin curve is found to lie well inside the envelope curve at 5 K. On the basis of these observations, we infer that the low-field step in *M(H)* well below $T_0$ is metamagnetic in its origin, indicating the existence of an antiferromagnetic component.

We now discuss the heat-capacity behavior of such a complex system. From figure 2g, it is clear that *C* shows a peak just below $T_o$ and the peak temperature is seen to increase as the magnetic field is increased. This was confirmed by magnetization data as well, measured in the presence of various magnetic fields (not shown here). Such a feature is characteristic of FM systems. There is a weak drop below 4 K, which could be attributed to the onset of the antiferromagnetic anomaly discussed above. The temperature response of isothermal entropy change ($\Delta S$ defined as $S(H)-S(0)$) is shown in figure 2h. These $\Delta S(T)$ curves were obtained from *M(H)* isotherms measured at different temperatures (under ZFC condition after cooling the sample from 50 K) employing the Maxwell's equation. Generally, for a FM system, $-\Delta S$ is expected to show a 'positive' peak at $T_0$ due to the loss of spin-disorder, while for an AFM a negative peak is observed [24]. The observation of the positive peak around $T_0$ (Fig 2h) is consistent with paramagnetic-ferromagnetic transition.

Temperature dependence of $\rho$ measured is shown in the inset of figure 3. A sharp change of slope is observed around $T_o$, which is smeared by the application of a magnetic field. The magnetoresistance [MR = $\{\rho(H)-\rho(0)\}/\rho(0)$] (in figure 3) shows interesting features. The sign of MR at 1.8 K is distinctly negative with a small magnitude for a wide range of magnetic fields and it becomes positive at a higher field resulting in a minimum in the plots of MR(*H*) at 15 kOe. The observed positive sign of MR at higher fields is generally uncharacteristic of FM and this could arise from the AFM component, proposed above. However, as the temperature is increased MR(*H*) curve lies in the negative quadrant for the entire field range of measurement. MR varies nearly as $H^{2/3}$, known for FM systems [25].

This study thus clearly establishes that this compound initially undergoes ferromagnetic ordering, confirming that this is unique among other members of $R_2PdSi_3$ series, apart from other low temperature anomalies. Clearly, origin of the magnetism of this compound is quite complex.

### B. 4f-hybridization in doped $Nd_2PdSi_3$

Now we focus on the modification of the magnetic properties of the title compound, based on substitutions at Nd site and at Si site, viz., on the series of $Nd_{2-x}(Y, La)_xPdSi_{3-y}Ge_y$ (x, y = 0.2, 0.4, 0.8, and 1.2). The lattice parameters tabulated in Table 1 reveal that, for *Y*-series the lattice constants *a* and *c* decrease with *x*, resulting in a decrease in unit cell volume (*V*), whereas for *La*-series, both increase leading to an increase in *V*. On the other hand, for the *Ge*-series, it is found that *a* increases but *c* decreases, however resulting in an increase in *V*.

Figures *4 a-c* show the temperature dependent magnetization measured in the presence of a magnetic field of 100 Oe for all the alloys. It is clear that $T_0$ (obtained from the peak of d(*M/H*)/d*T* and tabulated in Table 1) is suppressed for all the doped compounds. For the Y-series, $T_0$ is suppressed down to 2.9 K for the extreme composition (*x*= 1.2), while for La-series,

it is suppressed to 5.9 K only for the same level of doping. While the suppression of $T_0$ is a natural consequence of dilution of the magnetic sublattice, the difference in the magnitudes of suppression merit an attention. Interestingly, for Ge-series without disturbing Nd sublattice, a suppression of $T_0$ (to 7.5 K for $y=1.2$) is noted even for a marginal increase in volume. [Needless to state that this experimentally determined value of the effective moment obtained from the high temperature Curie-Weiss region is essentially around $3.5\mu_B/Nd^{3+}$, which is in accordance with free $Nd^{3+}$ value.]

In order to gain support for these trends in $T_o$, heat-capacity and electrical resistivity as a function of temperature were measured for all the compounds. The magnetic part ($C_m$) of $C$ was obtained by subtracting the lattice part, with $Y_2PdSi_3$ and $La_2PdSi_3$ as the reference for lattice contribution for the respective series employing a procedure given Ref 26. The trend observed in $T_0$, inferred from magnetization is also reflected in $C$ of all the series (figure *4 d-f*). The temperature dependent normalized resistivity $\rho(T)$ for all the members of the series is shown in the figure *4 g-i*. We have plotted normalized resistivity as the absolute values are not very reliable due to microcracks in the specimens. A change of slope in the curve is observed for all members near $T_0$ (also obtained from the peak of $d^2\rho/dT^2$ versus $T$ plots, not shown here). There is a good agreement in the trends of $T_0$ (within an error of $\pm 1$ K) obtained from magnetization, $C$ and $\rho$ measurements.

It is obvious from the above data that, though the trend in the unit cell volume variation is different for the three series, there is a suppression of $T_0$ in all cases. The magnitude of suppression is different among the three solid solutions. In order to quantitatively address these variations, the change in $T_o$ (i.e. $\Delta T_o$) normalized to Nd concentration is plotted in figure 5a as a function of the change in unit-cell volume ($\Delta V$) with respect to the parent compound. We first focus on the Y and La substitutions, both of which dilute the Nd-sublattice. For a given $\Delta V$, the change $\Delta T_o$ is negligible and nearly constant in La-series, which is consistent with expectations based on indirect exchange interaction. Interestingly, for Y-series, significant suppression of $T_o$ is observed. This suppression in Y-series compared to that in La-series points to strong 4f(*Nd*) hybridization with the valence electrons of neighboring ions. Since the substitution of Ge for Si results in an expansion of volume, just based on volume arguments alone, an enhancement in $\Delta T_o$ is expected on the basis of extrapolation of $\Delta V$-dependence observed in Y-series, in contrast to the present observations. In addition, for a given $\Delta V$, the degree of suppression of $T_o$ for Ge is much more than that for Y-series, emphasizing the role of 4f-ligand hybridization. Since we consider $\Delta T_o$ is a measure of the variation in the 4f-hybridization, then the hybridization is stronger for *Ge*-substitution compared to Y-substitution.

The conclusion that the ligand hybridization, rather than volume change, controls variations in magnetic behavior, is endorsed by the data obtained from high pressure studies. In figure 5b, the temperature dependencies of magnetization measured in a field of 100 Oe for various values of external pressure in the vicinity of magnetic transition are shown. It is obvious that $T_o$ remains essentially unchanged up to a pressure of 10 kbar. In order to make a comparison with the variations in the solid solutions, it is desirable to know the bulk modulus which is not available in the literature. It is known that the bulk modulus of Ce compounds falls in the range 60 – 100 GPa [27]. Though it is strictly not correct to assume that the bulk modulus of Nd compounds are the same as Ce systems, considering that 4f orbitals in Nd are relatively less hybridized compared to Ce-4f orbitals. In the absence of any knowledge of bulk modulus at present, we use these values for a rough idea, as the above range of bulk modulus is sufficiently wide. With this caution, for the present alloys, one should have seen an observable variation of $T_0$ under pressure. For instance, the pressure exerted by Y substitution for $x= 0.3$ is of the order of 10 kbar, assuming bulk modulus of 100 GPa. Clearly, $V$ alone cannot account for the changes in $T_o$ in solid solutions.

It should however be noted (see table 1) that there is a contraction along $c$-axis with gradual substitution in Y and Ge-series, whereas in the La-series, there is an expansion along $c$-direction. The Pd-Si(Ge) layers (which are sandwitched between Nd layers and) stacked along $c$-direction come closer to Nd ions in Y and Ge systems unlike in La substituted alloys, thereby facilitating 4f-hybridization with the ligand orbitals in the former cases. Qualitatively speaking, this appears to favor our conclusion of enhancement of 4f-hybridization in Y and Ge based solid solutions. As the sign of $\theta_p$ is positive along $c$-direction as inferred from the data on single crystals [21], the compression and enhanced hybridization along this direction may be directly related to depression of ferromagnetic interaction in these alloys. However, relatively smaller values of $\Delta T_0$ are observed for Y-series while compared to Ge-series for a given $\Delta c$ without a quantitative correlation between $\Delta c$ and $\Delta T_0$. This discrepancy can be reconciled by proposing that, in the Y-series, the changes induced by the suppression of $a$-parameter has counterbalancing effect on $\Delta T_o$.

With the scenario discussed above, with respect to the differences in 4f-hybridization in the solid solutions, it is of interest to compare the ac-$\chi$ behavior. Real part of ac-$\chi$ at different frequencies is plotted as a function of temperature for the parent compound and the end members in figure 6. It is obvious that ac-$\chi$ is frequency-dependent below $T_o$ for all compositions, but the magnitude of the shift in peak temperature (say, for a variation of frequency from 1.3 to 931 Hz) is more prominent for Y and Ge substituted samples. The peak shift for the La alloy could not be resolved at all. These findings signal that an increase in 4f-hybridization favors glassy features. The shift of the peak temperature with frequency is fitted to power law (which follows from dynamic scaling theory) which is of the form [28]

$$\tau = \tau^0 (T/T_f-1)^{-z\nu}$$

where, $\tau^0$ is the microscopic flipping time, $T_f$ is the glass-freezing temperature, $\nu$ is the critical exponent which describes the growth of spin correlation length and z is the dynamic exponent which describes the slowing down of relaxation [Insets of Fig. 6]. For the Y alloy, the fit yielded $z\nu = 18 \pm 0.35$, $\tau^0 = 4*10^{-8}$ and $T_f = 1.6$ K. For $Nd_2PdSi_{1.8}Ge_{1.2}$, the shift in peak temperature when fitted with power law yields $z\nu = 3 \pm 0.2$, $\tau^0 = 10^{-9}$ and $T_f = 8.5$ K. Characterization of the frequency dependence of $T_f$ was carried out by relative shift in $T_f$ per decade of frequency i.e. $\delta f = \Delta T_f / T_f * \Delta \ln \nu$. Even though the values of $\delta f$ lies in the range of 0.03-0.003, which is similar to that observed for spin-glasses [29], we do not classify these compounds as a spin-glass considering that the features in heat-capacity are well-defined at the magnetic transition.

## IV. Conclusions

The present study establishes that magnetism of $Nd_2PdSi_3$ is unique in the sense that this compound orders ferromagnetically (~ 15.5 K) unlike other members of this series, in addition to possible onset of antiferromagnetic correlations at much lower temperatures. We have identified a Nd-based compound, in which Nd 4f hybridization apparently plays a major role on the magnetism, making it different from other members in this ternary family. This conclusion is based on a comparative investigation of the solid solutions, $Nd_{2-x}(Y, La)_xPdSi_{3-y}Ge_y$ (x, y = 0.2, 0.4, 0.8, and 1.2) as well as high pressure studies. These studies reveal a strengthening 4f($Nd$) hybridization with the valence electrons of other ions in Y or Ge substituted samples. Interestingly, stronger hybridization seems to favor development of glassy features near the magnetic transitions. In short, this work emphasizing the role of Nd 4f hybridization on magnetism adds a new input to the field of strongly correlated electron systems. We believe that such systems could open up an exciting avenue to understand heavy-fermion behavior and quantum critical point effects among Nd-based systems, for instance, by extending studies to very high pressures.

**Acknowledgement**
The authors would like to thank W. Loeser and Y. Xu (IFW, Dresden) for making their single crystals available to us for some magnetization measurements.

*sampath@mailhost.tifr.res.in

**Table 1.** The lattice constants ($a, c, \pm 0.004$ Å), unit-cell volume ($V$) and the magnetic ordering temperature ($T_0 \pm 0.5$ K) for the alloys, $Nd_{2-x}(Y, La)_x PdSi_{3-y}Ge_y$ ($x, y$ = 0.2, 0.4, 0.8, and 1.2).

| $x, y$ | $a$(Å) | $c$(Å) | $V$(Å)$^3$ | $T_0$(K) |
|---|---|---|---|---|
| 0.0 | 8.216 | 8.414 | 491.9 | 15.5 |
| $Y_x$ = 0.2 | 8.202 | 8.394 | 489.1 | 13.4 |
| 0.4 | 8.187 | 8.348 | 484.5 | 10.0 |
| 0.8 | 8.164 | 8.311 | 479.8 | 5.2 |
| 1.2 | 8.147 | 8.227 | 472.9 | 2.9 |
| $La_x$=0.2 | 8.228 | 8.443 | 495.1 | 14.3 |
| 0.4 | 8.229 | 8.479 | 497.3 | 13.0 |
| 0.8 | 8.252 | 8.545 | 503.9 | 9.9 |
| 1.2 | 8.273 | 8.607 | 510.2 | 5.9 |
| $Ge_y$=0.2 | 8.222 | 8.414 | 492.6 | 13.0 |
| 0.4 | 8.235 | 8.412 | 494.1 | 12.5 |
| 0.8 | 8.274 | 8.405 | 498.2 | 10.7 |
| 1.2 | 8.284 | 8.388 | 499.7 | 7.5 |

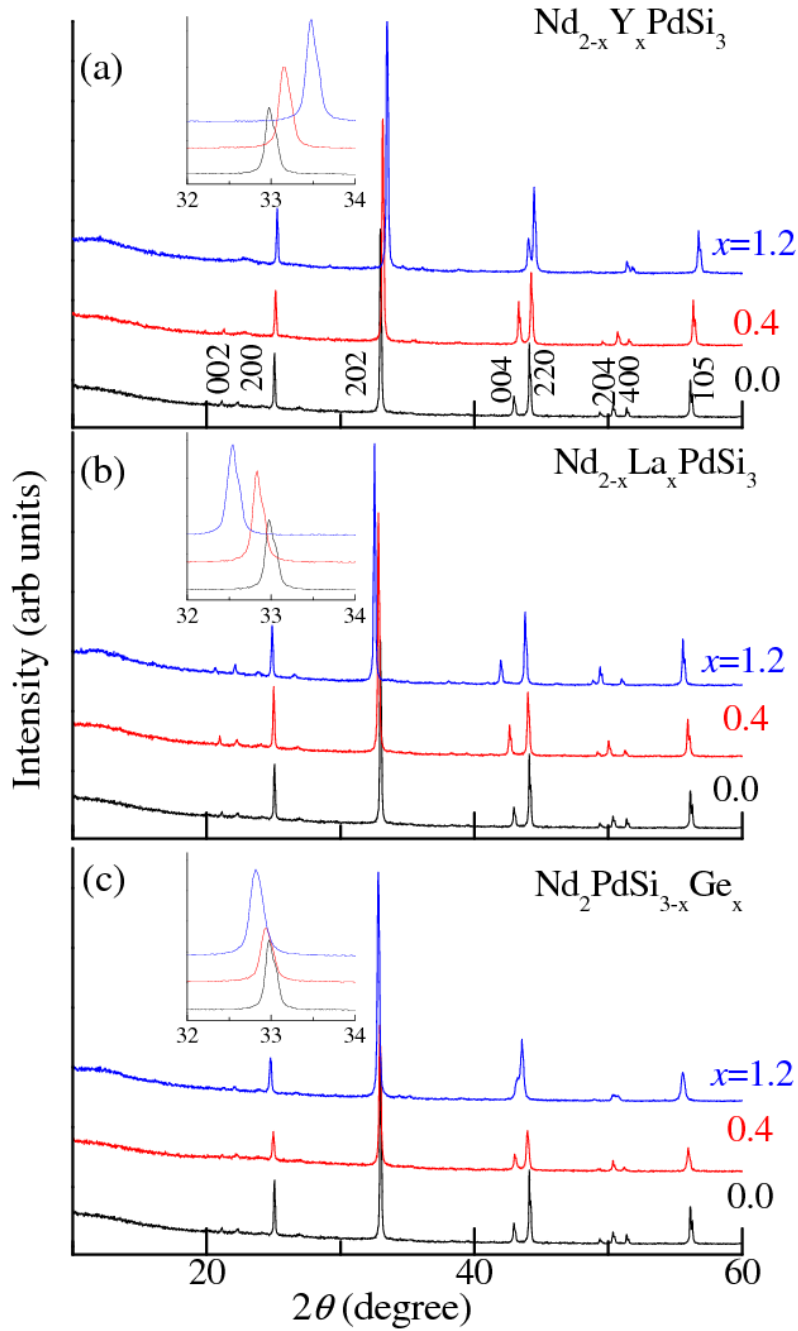

Figure 1: (color online) X-ray diffraction patterns (Cu $K_\alpha$) below $2\theta=60^o$ for the alloys, $Nd_{2-x}(Y, La)_xPdSi_{3-y}Ge_y$ (for selected compositions). The curves are shifted along y axis for the sake of clarity. Inset: The region around the main peaks is shown in an expanded form to show a gradual shift of diffraction lines with changing composition.

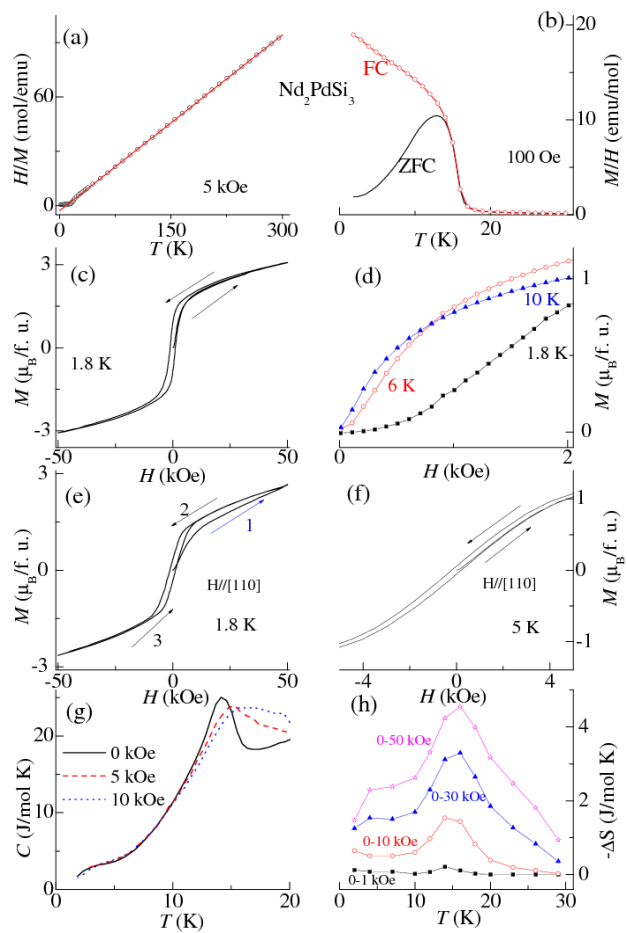

Figure 2: (color online) For polycrystalline Nd$_2$PdSi$_3$, (a) temperature (*T*) dependence of inverse susceptibility with the line through the data points showing the Curie-Weiss fit to the susceptibility, (b) *T*-dependence (1.8-30 K) of magnetization (*M*) divided by magnetic field (*H*) for zero-field-cooled (ZFC) and field cooled (FC) conditions, (c) magnetic hysteresis behavior at 1.8 K, and (d) isothermal magnetization behavior at different temperatures below H≤ 2 kOe. For single crystalline sample, magnetic hysteresis behavior (data taken in the range 50 to -50 kOe) at 1.8 and 5 K are shown in (e) and (f) for H//[110]. (g) Heat capacity as a function of temperature in the presence of different magnetic fields, and (h) isothermal entropy change (ΔS) as a function of temperature for magnetic fields up to 50 kOe for the polycrystal. Unless stated, the lines through the data points serve as guides to the eyes.

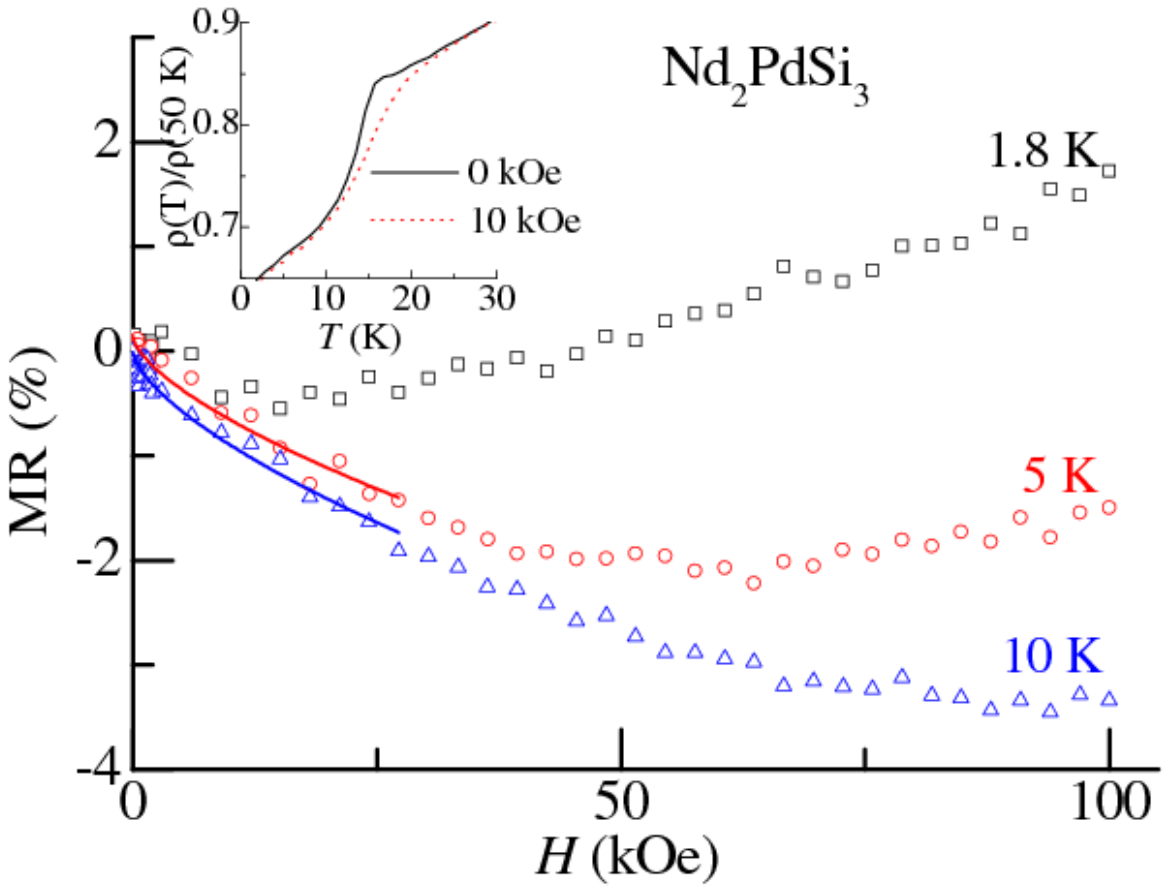

Figure 3: (color online) (a) Magnetoresistance (MR) as a function of external magnetic field at selected temperatures. The continuous lines through 5 and 10 K data are drawn to show that MR varies as $H^{2/3}$ at low fields. Inset: Electrical resistivity ($\rho$) as a function of temperature (1.8-30 K).

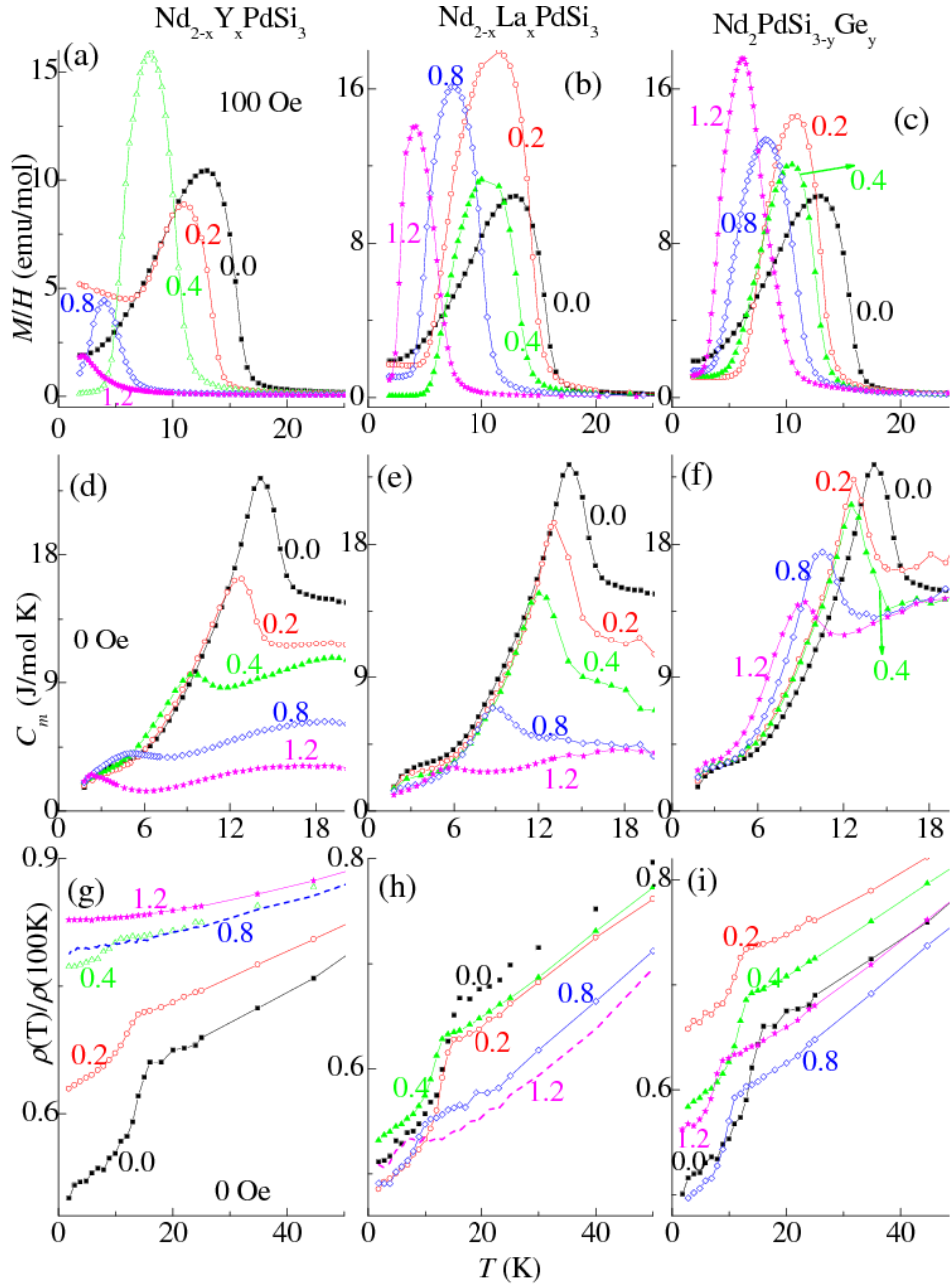

Figure 4 (color online) (a-c) Temperature ($T$) dependence (1.8-30 K) of magnetization ($M$) divided by magnetic field ($H$) for $Nd_{2-x}(Y, La)_xPdSi_{3-y}Ge_y$ (x, y = 0.2, 0.4, 0.8, and 1.2) for zero-field-cooled condition of the specimen. (d-f) Magnetic specific heat ($C_m$) as a function of temperature in zero field and (g-i) Normalized electrical resistivity as a function of temperature in zero field for all the compounds.

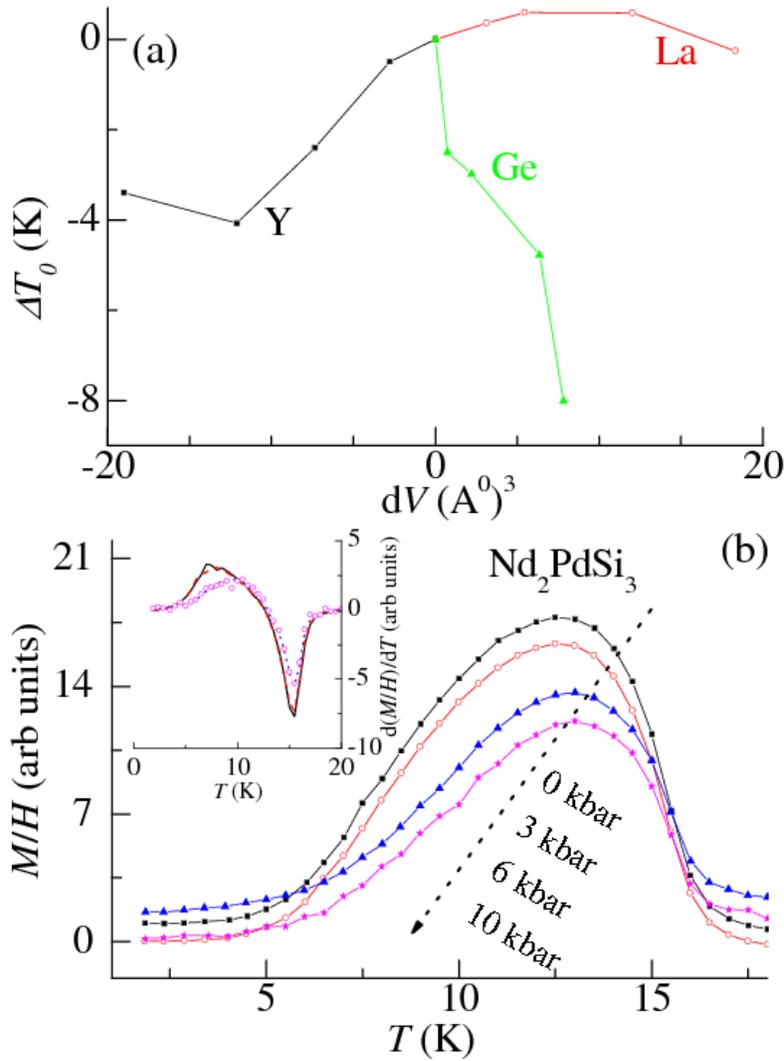

Figure 5 (color online): (a) The change $\Delta T_o$ in magnetic ordering temperature $T_o$ normalized to Nd concentration plotted as a function of change in unit-cell volume ($\Delta V$) with respect to the parent compound, for the compounds in the series, $Nd_{2-x}(Y, La)_xPdSi_{3-y}Ge_y$ (x, y = 0.2, 0.4, 0.8, and 1.2). (b) Magnetization divided by magnetic field obtained in a field of 100 Oe under the influence of external pressure. The curves are shifted along *y*-axis for the sake of clarity. Inset: Temperature response of d(*M*/*H*)/dT for all external pressures to show that the magnetic ordering temperature does not vary with pressure.

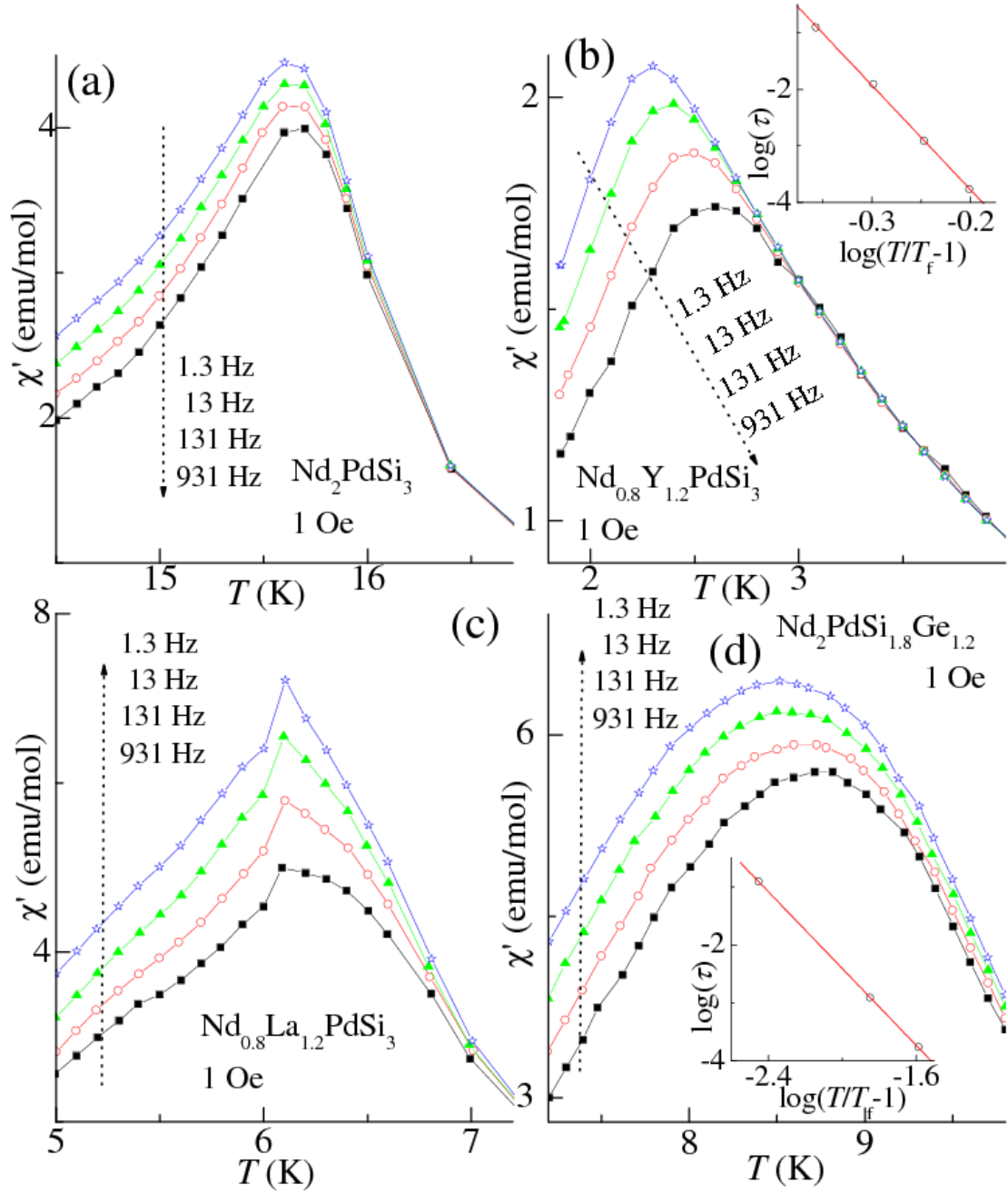

Figure 6 (color online): Real part of ac susceptibility at different frequencies plotted as a function of frequency for (a) $Nd_2PdSi_3$, (b) $Nd_{0.8}Y_{1.2}PdSi_3$, (c) $Nd_{0.8}La_{1.2}PdSi_3$ and (d) $Nd_2PdSi_{31.8}Ge_{1.2}$. Insets show dynamical scaling fit of peak temperature with reduced temperature for $Nd_{0.8}Y_{1.2}PdSi_3$ and $Nd_2PdSi_{1.8}Ge_{1.2}$.